\documentclass[amssymbols,12pt]{article}
  \usepackage{amssymb}
  \usepackage{graphicx}
  \usepackage{psfrag}
  \usepackage{amsmath}

\newcommand{\curly}[2]
{\left( \begin{array}{c} #1 \\ #2 \end{array} \right)}

\newcommand{\curl}[4]
{\left( \begin{array}{cc} #1 & #2 \\[2ex] #3 & #4 \end{array}
\right)}

\newcommand\U{ {\mathcal{U}} }
\newcommand\V{ {\mathcal{V}} }
\newcommand\diag{ {\mathrm{diag}} }

\newcommand\LL{ {\mathcal{L}} }
\newcommand\MM{ {\mathcal{M}} }
\newcommand\be{\begin{eqnarray}}
\newcommand\barray{$$\begin{array}{rl}}
\newcommand\ee{\end{eqnarray}}
\newcommand\earray{\end{array}$$}

\newcommand\half{\frac{1}{2}}

\newcommand\quarter{\frac{1}{4}}

\newcommand\s{Schr\"odinger }

\DeclareMathOperator*{\Res}{Res}

\begin{document}
\title{Linearizable Initial-Boundary Value Problems for the \\
sine-Gordon Equation on the Half-Line}
\author{A.S. Fokas  \\
{\em Department of Applied Mathematics} \\
{\em and Theoretical Physics } \\
{\em University of Cambridge } \\
{\em Cambridge, CB30WA, UK} \\
{\em t.fokas@damtp.ac.uk} \\
\\
Dedicated to I.M. Gelfand \\
on the occasion of his ninetieth birthday}

\date{September  2003}

\maketitle

\begin{abstract}
A rigorous methodology for the analysis of initial boundary value problems on the half-line,
$0<x<\infty$, $t>0$, for integrable nonlinear evolution PDEs has recently appeared in the
literature.  As an application of this methodology the solution $q(x,t)$ of the sine-Gordon
equation can be obtained in terms of the solution of a $2\times 2$ matrix Riemann-Hilbert
problem.  This problem is formulated in the complex $k$-plane and is uniquely defined in terms
of the so called spectral functions $a(k)$, $b(k)$, and $B(k)/A(k)$.  The functions $a(k)$ and $b(k)$ can
be constructed in terms of the given initial conditions $q(x,0)$ and $q_t(x,0)$ via the
solution of a system  of two {\it  linear} ODE's, while for \emph{arbitrary} boundary conditions the
functions $A(k)$ and $B(k)$ can be constructed in terms of the given boundary condition via the
solution of a system of four {\it nonlinear} ODEs.  In this paper we analyse two \emph{particular}
boundary conditions: the case of constant Dirichlet data, $q(0,t) = \chi$, as well as the case
that $q_x(0,t)$, $\sin (q(0,t)/2)$, and $\cos(q(0,t)/2)$ are linearly related by two constants
$\chi_1$ and $\chi_2$.  We show that for these particular cases, the system of the above
nonlinear ODEs can be avoided, and $B(k)/A(k)$ can be computed explicitly in terms of $\{
a(k),b(k),\chi\}$ and $\{ a(k),b(k), \chi_1, \chi_2\}$ respectively.  Thus these ``linearizable''
initial-boundary value problems can be solved with absolutely the same level of efficiency as
the classical initial value problem of the line.
\end{abstract}

\section{Introduction}
Let the real function $q(x,t)$ satisfy an initial-boundary value problem for the
sine-Gordon equation on the half-line, $\{ 0<x<\infty, 0<t<T\}$, where $T$ is a
positive constant.  The function $q(x,t)$ can be constructed as follows [1],[2]:
\begin{itemize}
\item Given initial conditions construct the spectral functions $a(k)$ and $b(k)$.
These functions are defined in terms of $\phi(x,k)$, where the vector $\phi$ is an
appropriate solution of the $x$-part of the associated Lax pair evaluated at $t=0$.
Thus $\phi$ is defined in terms of the initial conditions $q(x,0)$ and $q_t(x,0)$. 
\item From the given boundary condition, characterize the unknown boundary value at
$x=0$ by the requirement that the spectral functions $\{ a(k), b(k), A(k), B(k)\}$
satisfy the global relation
$$ 
a(k)B(k) - b(k)A(k) = e^{\frac{i}{4}(k+\frac{1}{k})T} c(k), \quad
{\mathrm{Im}}\: k \geq 0, \quad k \neq 0, \eqno (1.1)
$$
where $c(k)$ is analytic for ${\mathrm{Im}}\: k >0$ and is of $O(1/k)$ as $k\rightarrow\infty$.
The functions $A(k)$ and $B(k)$ are defined in terms of $\Phi(t,k)$, where the vector
$\Phi$ is an appropriate solution of the $t$-part of the associated Lax pair evaluated
at $x=0$.  Thus $\Phi$ is defined in terms of $q(0,t)$ and $q_x(0,t)$. 
\item Given $\{ a(k),b(k),B(k)/A(k)\}$, construct $q(x,t)$ through the solution of a
$2\times 2$ matrix Riemann-Hilbert problem.  The function $q(x,t)$ satisfies the
sine-Gordon equation as well as the given initial and boundary conditions.
\end{itemize}

The most complicated step in the above construction is the characterization of the
missing boundary value.  For example, for the Dirichlet problem where the function
$q(0,t)$ is prescribed as the boundary condition, it is shown in [2] that the unknown
boundary value $q_x(0,t)$ can be obtained through the solution of a system of four {\it
nonlinear} ODEs.

It was shown in [1] and [3] that for some particular boundary conditions, which we
refer to as \emph{linearizable} boundary conditions, it is possible to bypass the above system
of nonlinear ODEs and to construct $B(k)/A(k)$ using only algebraic manipulations.  In
particular, it was shown in [1] that this is the case for the boundary condition
$q(0,t) = \chi$, $\chi$ constant.

In this paper we show that there exists another linearizable boundary condition
which involves two constants $\chi_1$ and $\chi_2$.  For completeness we also include the
case $q(0,t) = \chi$.

\paragraph{Theorem 1.1}  Let the real function $q(x,t)$ satisfy the sine-Gordon
equation
$$ q_{tt} - q_{xx} + \sin q=0, \quad 0<x<\infty, \quad 0< t<T, \eqno (1.2)$$
(where $T$ is a given constant)  the initial conditions
$$ q(x,0) = q_0(x), \quad q_t(x,0) = q_1(x), \quad 0<x<\infty, \eqno (1.3)$$ 
and either of the following two boundary conditions, 
$$ q(0,t) = \chi, \eqno (1.4a)$$
or
$$ q_x(0,t) + \chi_1 \cos \left( \frac{q(0,t)}{2}\right) + \chi_2\sin \left(
\frac{q(0,t)}{2}\right) = 0, \eqno (1.4b)$$
where $\chi,\chi_1,\chi_2$ are real constants.  Assume that $q_0(x) -2\pi m$ and
$q_1(x)$ are Schwartz functions for $m$ integer, and that the initial conditions are
compatible with the boundary conditions at $x=t=0$, i.e.~assume that for (1.4a) and
(1.4b) the following conditions are valid respectively
$$q_0(0)=\chi \ \ {\mathrm{and}} \ \ q_1(0) =0, \quad \dot q_0(0) + \chi_1\cos \left(
\frac{q_0(0)}{2}\right) + \chi_2\sin \left( \frac{q_0(0)}{2}\right) =0.$$
The above initial-boundary value problems have a unique global solution given by
$$ \cos q(x,t) = 1+2\lim_{k\rightarrow \infty} \left\{ (k\mu_{12}(x,t,k))^2 +
2i\partial_x(k\mu_{22}(x,t,k))\right\}, \eqno (1.5)$$
where $\mu_{12}$ and $\mu_{22}$ denote the (12) and (22) entries of the $2\times 2$
matrix $\mu(x,t,k)$ which satisfies the following Riemann-Hilbert problem.

\begin{figure}[h] 
\psfrag{a}{$D_{1}$}
\psfrag{b}{$D_{2}$}
\psfrag{c}{$D_{3}$}
\psfrag{d}{$D_{4}$}
\begin{center}
\includegraphics{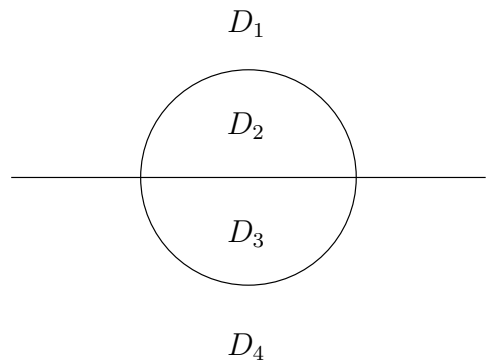}
\end{center}
\caption{The contour $\LL$ and the domains
  $D_j,j=1,\cdots,4.$}\label{fig1.1}
\end{figure}

(i) $\mu$ is meromorphic in $k$ for $k \in {\mathbb{C}} \backslash \LL$, where $\LL$
, which is depicted in Figure~\ref{fig1.1}, is defined by 
$$
\LL = \{ \mathrm{Re}\; k =0 \cup |k|=1\}.
$$
\vskip .2in
(ii) Let the domains $D_j, j=1,\cdots,4$ which are depicted in Figure~\ref{fig1.1}, be defined by
$$ D_1 = \{ {\mathrm{Im}} \ \ k >0 \cap |k|>1\}, \quad D_2 = \{ {\mathrm{Im}}\: k>0 \cap |k|<1\}, $$
$$ D_3 = \{ {\mathrm{Im}} \ \ k <0 \cap |k|<1\}, \quad D_4 = \{ {\mathrm{Im}}\: k<0 \cap |k|>1\}. $$
The matrix $\mu$ satisfies the jump condition
$$\mu_-(x,t,k) = \mu_+(x,t,k) J(x,t,k), \quad k \in \LL, \eqno (1.6)$$
where $\mu$ is $\mu_-$ for $k\in D_2\cup D_4$, $\mu$ is $\mu_+$ for $k\in D_1\cup
D_3$, and the $2\times 2$ matrix $J$ is defined in terms of the spectral functions $\{
a(k),b(k), B(k)/A(k)\}$ and the explicit function $\theta(x,t,k)$ by the following
formulae:
\begin{align*}
J &=J_1, \quad k\in D_1\cap D_2;& J &=J_2, \quad k\in D_2\cap D_3;& \\
J &=J_3, \quad k\in D_3\cap D_4;& J &=J_4, \quad k\in D_4\cap D_1,
\end{align*}
\begin{figure}[h] 
\psfrag{a}{$J_{1}$}
\psfrag{b}{$J_{2}$}
\psfrag{c}{$J_{3}$}
\psfrag{d}{$J_{4}$}
\begin{center}
\includegraphics{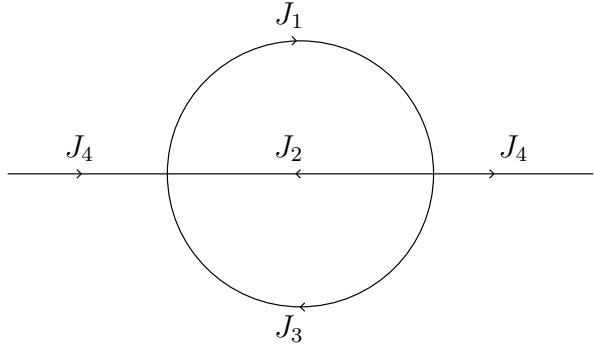}
\end{center}
\caption{The jump matrix $J$}\label{fig1.2}
\end{figure}
\vskip .2in
$$J_1 = \curl{1}{0}{\Gamma(k)e^{2i\theta}}{1}, \quad J_3 = \curl{1}{\overline{\Gamma(\bar
k)}e^{-2i\theta}}{0}{1}, $$
$$ J_4=\curl{1}{-\gamma(k)e^{-2i\theta}}{-\bar \gamma(k)e^{2i\theta}}{1 +|\gamma(k)|^2}, \quad
J_2 = J_3J_4^{-1}J_1,$$
$$ \gamma(k) = \frac{b(k)}{\overline{a(k)}}, k\in {\mathbb{R}}; \quad \Gamma(k) = - \frac{
\frac{\overline{B(\bar k)}}{\overline{A(\bar k)}}}{a(k) \left[ a(k) + b(k)
\frac{\overline{B(\bar k)}}{\overline{A(\bar k)}}\right]}, \quad k \in D_2; \eqno
(1.7)$$
$$\theta(x,t,k) = \quarter(k - \frac{1}{k})x + \quarter(k + \frac{1}{k})t. $$
The functions $a(k)$ and $b(k)$ are defined in terms of $q_0(x)$ and $q_1(x)$ by
$$ a(k) = \phi_2(0,k), \quad b(k) = \phi_1(0,k), \quad {\mathrm{Im}} \ \ k \geq 0, \eqno (1.8)$$
where the vector $\phi(x,k)$ with component $\phi_1(x,k)$ and $\phi_2(x,k)$ satisfies
$$ 
\phi_x + \frac{i}{4}\left(k-\frac{1}{k}\right) \sigma_3\phi = \quarter \curl{i(k+ \frac{\cos q_0(x)}{k})}{-i (\dot q_0(x)
+ q_1(x)) - \frac{\sin q_0(x)}{k}}{-i (\dot q_0(x) + q_1(x)) + \frac{\sin q_0(x)}{k}}{-
i( k+ \frac{\cos q_0(x)}{k})} \phi, 
$$
$$ 0<x<\infty, \ \  {\mathrm{Im}} \ \ k \geq 0,$$
$$\phi(x,k) = e^{ikx} \left( \curly{0}{1} + o(1)\right) \ \ {\mathrm{as}} \ \ x
\rightarrow \infty, \eqno (1.9)$$
and $\sigma_3 = $ diag $(1,-1)$.  

For $k \in D_{1}$ the ratio $B(k)/A(k)$ equals $b(k)/a(k)$.  For $k
\in D_3$, the ratio $B(k)/A(k)$ for the cases (1.4a) and (1.4b) is given
respectively by
$$\frac{B(k)}{A(k)} = \frac{f(k)b(\frac{1}{k}) - a( \frac{1}{k})}{f(k)a(\frac{1}{k}) -
b(\frac{1}{k})}, \eqno (1.10a)$$
and 
$$ \frac{B(k)}{A(k)} = \frac{b(\frac{1}{k}) [\alpha(k) +
    \overline{\alpha(\bar{k})} + 2i\cos{(\frac{q_{0}(0)}{2})}] +
    a(\frac{1}{k}) [ \overline{\alpha(\bar{k})} - \alpha(k) -
    2\sin{(\frac{q_{0}(0)}{2})}]}{ a(\frac{1}{k}) [\alpha(k) +
    \overline{\alpha(\bar{k})} - 2i\cos{(\frac{q_{0}(0)}{2})}] -
    b(\frac{1}{k}) [ \alpha(k) - \overline{\alpha(\bar{k})} -
    2\sin{(\frac{q_{0}(0)}{2})}]} \eqno (1.10b)$$
where
$$
f(k) = i \frac{k^2+1}{k^2-1} \frac{\sin \chi}{\cos{\chi}-1}, \quad
\alpha(k) = \frac{i\chi_{1}}{k + \frac{1}{k}} + \frac{\chi_{2}}{k -
  \frac{1}{k}}. \eqno (1.11)
$$

(iii) Define the function $\Delta(k)$ by
$$
\Delta(k) = a(k) \overline{D(\bar{k})} + b(k) \overline{N(\bar{k})},
\eqno (1.12),
$$
where $D(k)$ and $N(k)$ denote the denominator and the numerator of
the rhs of equations (1.10).  If the function $a(k)$ has zeros for Im
$k >0$, and/or the function $\Delta(k)$ has zeros in $D_2$, then
$\mu(x,t,k)$ satisfies appropriate residue conditions, see section 4.

\paragraph{Organization of the Paper}

In section 2 we summarize the methodology of [1], [3] for identifying and analyzing
linearizable boundary conditions. If a given PDE admits different Lax
pair [4] formulations, it is possible to search for linearizable boundary conditions for each of
these different formulations. The Lax  pair analyzed in [1], see equations (2.9), gives rise to
the case (1.4a); the sG also admits an alternative Lax pair [5], see (2.11), which gives rise to
the linearizable case (1.4b).  Since the basic RH problem presented in Theorem 1.1 is associated
with the Lax pair analyzed  in [1], we present in section 3 a general formalism which connects the
spectral functions $\{ A(k),B(k) \}$ associated with two different Lax pairs.  As an application
of this formalism we use the alternative Lax pair (2.11) to identify (1.4b), but we solve the sG,
for both boundary conditions (1.4a) and (1.4b), using the Lax pair of
[1].  In section 4 we derive the residue conditions and in section 5
we discuss further these results.

\section{An Overview, Different Lax Pairs for the sine-Gordon, and Linearizable Boundary
Conditions}
We first recall the definition of the spectral functions $A(k)$ and $B(k)$ [1]. 

\paragraph{Definition 2.1}  Suppose that the $t$-part of the Lax pair of a given integrable PDE is
$$ \Psi_t + if_2(k)\sigma_3\Psi = \tilde Q(x,t,k) \Psi, \eqno (2.1)$$
where 
$$ \sigma_3 = \diag(1,-1), \eqno (2.2)$$
$\Psi(x,t,k)$ is a $2\times 2$ matrix, the scalar $f_2(k)$ is an analytic function of $k$, and the
$2\times 2$ matrix $\tilde Q(x,t,k)$ is an analytic function of $k$, of $q(x,t)$, of $\bar
q(x,t)$, and of the derivatives of these functions. Let $M(t,k)$ be the unique solution of
$$M_t + if_2(k)\sigma_3M = \tilde Q(0,t,k)M, \quad 0<t<T, \quad k\in {\mathbb{C}}, $$
$$ M(0,k) = \diag (1,1), \eqno (2.3)$$
where $T$ is a positive constant.  Assume that $\tilde Q(0,t,k)$ is such that $M(t,k)$ has the
form
$$M(t,k) = \curl{\overline{\Phi_2(t,\bar k)}}{\Phi_1(t,k)}{\rho\overline{\Phi_1(t,\bar
k)}}{\Phi_2(t,k)}, \quad \rho^2=1. \eqno (2.4)$$
The spectral functions $A(k)$ and $B(k)$ are defined by
$$ A(k) = e^{if_2(k)T} \overline{\Phi_2(T,\bar k)}, \quad B(k) = -e^{if_2(k)T} \Phi_1(T,k), \quad
k \in {\mathbb{C}}. \eqno (2.5)$$

\paragraph{Linearizable Boundary Conditions} \ \ 

Let the transformation $k\rightarrow \nu(k)$ be defined by the requirement that it leaves $f_2(k)$
invariant, i.e.
$$ f_2(\nu(k)) = f_2(k), \quad k \in {\mathbb{C}}. \eqno (2.6)$$
Suppose that there exists a nonsingular $2\times 2$ matrix $N(k)$ such that
$$ N(k)^{-1} \left[ if_2(k)\sigma_3 - \tilde Q(0,t,\nu(k))\right] N(k) = if_2(k)\sigma_3-\tilde
Q(0,t,k). \eqno (2.7)$$
Then the solution $M(t,k)$ of equations (2.3) satisfies
$$M(t,\nu(k)) = N(k)M(t,k)N(k)^{-1}. \eqno (2.8)$$
This equation and the definitions (2.5), imply a relation between $\{ A(\nu(k)), B(\nu(k))\}$ and
$\{ A(k),B(k)\}$.  Using this relation, it is possible to compute $B(k)/A(k)$ using only the
algebraic manipulation of the global relation [1].

\paragraph{Two Different Lax Pairs for the sine-Gordon} \ \ 

The sG possesses the Lax pair [6]
$$ \Psi_x + \frac{i}{4}(k - \frac{1}{k})\sigma_3\Psi = Q(x,t,k)\Psi, $$
$$ \Psi_t + \frac{i}{4}(k + \frac{1}{k})\sigma_3\Psi = \tilde
Q(x,t,k)\Psi, \quad k \in {\mathbb{C}}, \quad k \neq 0,
\eqno (2.9)$$
where $\Psi(x,t,k)$ is a $2\times 2$ matrix, $\tilde Q(x,t,k) = Q(x,t,-k)$, and the $2\times 2$
matrix $Q(x,t,k)$ is defined by
$$Q(x,t,k) = \quarter \curl{ \frac{i}{k}(-1+\cos q)}{-i (q_x + q_t) - \frac{\sin q}{k}}{-i (q_x +
q_t) + \frac{\sin q}{k}}{- \frac{i}{k}(-1 + \cos q)} \eqno (2.10)$$
with $q$ denoting $q(x,t)$. 

The sG also possesses the alternative Lax pair [5]
$$ \Psi_x = \U(x,t,k)\Psi, \quad \Psi_t = \V(x,t,k)\Psi, \quad k \in
{\mathbb{C}}, \quad k \neq 0, \eqno (2.11)$$
where $\Psi(x,t,k)$ is a $2\times 2$ matrix and the $2\times 2$ matrices $\U(x,t,k)$ and
$\V(x,t,k)$ are defined by
$$ \U = \quarter \left( \begin{array}{cc}
-iq_t & ke^{\frac{iq}{2}} - \frac{1}{k}e^{-\frac{iq}{2}} \\ \\ 
-ke^{- \frac{iq}{2}} + \frac{1}{k}e^{\frac{iq}{2}} & iq_t \end{array}\right), \quad \V = \quarter
\left( \begin{array}{cc}
-iq_x & ke^{\frac{iq}{2}} + \frac{1}{k}e^{-\frac{iq}{2}} \\ \\
-ke^{- \frac{iq}{2}} - \frac{1}{k}e^{\frac{iq}{2}} & iq_x \end{array} \right). \eqno (2.12)$$
\paragraph{Linearizable Case of the Second Lax Pair}
We now show that by analyzing the $t$-part of the second Lax pair (2.11) evaluated at $x=0$, it is
possible to identify linearizable boundary conditions using the simple formulation reviewed
earlier, see equations (2.6)-(2.8).

\paragraph{Proposition 2.1}  Let the $2\times 2 $ matrix $\MM(t,k)$ be the following solution of
the $t$-part of the Lax pair (2.11) evaluated at $x=0$:
$$\MM_t = \V(t,k)\MM, \quad 0<t<T, \quad k \in {\mathbb{C}}, $$
$$ \MM (0,k) = \diag(1,1), \eqno (2.13)$$
where
$\V(t,k) = \V(0,t,k)$ and $\V(x,t,k)$ is defined by equation (2.12b).  Then $\MM$ satisfies the
``symmetry'' relation
$$ \MM(t, \frac{1}{k}) = N(k)\MM(t,k)N(k)^{-1}, \quad 0<t<T, \quad k
\in {\mathbb{C}}, \quad k \neq 0, \eqno (2.14)$$
where for equations (1.4a) and (1.4b), $N(k)$ is given respectively by 
$$N(k) = \curl{1}{0}{0}{\frac{ e^{-\frac{i\chi}{2}} + k^2e^{\frac{i\chi}{2}} }{e^{\frac{i\chi}{2}}
+ k^2e^{- \frac{i\chi}{2}}}} \eqno (2.15)$$
and 
$$N(k) = \curl{ \frac{i\chi_1}{k+\frac{1}{k}} + \frac{\chi_2}{k-\frac{1}{k}} }{1}{-1}{
\frac{\chi_2}{k-\frac{1}{k}} - \frac{i\chi_1}{k+\frac{1}{k}}} . \eqno (2.16)$$

\noindent{\it Proof} In this case $f_2(k) = (k+1/k)/4$, thus equation (2.6) implies $\nu =1/k$.
Let us introduce the following notations for the matrix $N(k)$, 
$$N_{11} =N_1, \quad N_{12} = N_2, \quad N_{21} = N_3, \quad N_{22} = N_4. \eqno (2.17)$$
In this case equation (2.8) is
$$\V(t,\nu(k)) = N(k)\V(t,k)N(k)^{-1}. \eqno (2.18)$$
If $N_3=-N_2$, then the (11) and (22) entries of this equation are identically satisfied, while
both the (12) and (21) entries of equation (2.18) yield the equation
$$ 2iq_x(0,t)N_2 + e^{- \frac{iq}{2}(0,t)} \left( \frac{N_1}{k} - kN_4 \right) +
e^{\frac{iq}{2}(0,t)} \left(kN_1 - \frac{N_4}{k}\right) =0. \eqno (2.19)$$

We distinguish two cases

(i) $N_2=0, \quad N_4 = \alpha(k)N_1.$

If $\alpha(k)$ equals the (22) entry of the matrix $N$ defined in (2.15) then equation (2.19)
becomes the boundary condition (1.4a).

(ii) $N_2 \neq 0$, $N_1 =\alpha(k)N_2$, $N_4 = \beta(k)N_2$.

Then 
$$ q_x(0,t) + \frac{i}{2}(\beta-\alpha)(k + \frac{1}{k})\cos{\left(\frac{q(0,t)}{2}\right)} + \half
(\beta+\alpha)(k - \frac{1}{k}) \sin{\left(\frac{q(0,t)}{2}\right)} =0. \eqno (2.20)$$
Letting 
$$ \frac{i}{2}(\beta-\alpha) (k+\frac{1}{k}) = \chi_1, \quad \half (\beta+\alpha) (k -
\frac{1}{k}) = \chi_2,$$
i.e.
$$ \alpha(k) = \frac{i\chi_1}{k+\frac{1}{k}} + \frac{\chi_2}{k-\frac{1}{k}}, \quad \beta(k) =
\frac{\chi_2}{k-\frac{1}{k}} - \frac{i\chi_1}{k+\frac{1}{k}} = \overline{\alpha(\bar{k})}, \eqno (2.21)$$
equation (2.20) becomes the boundary condition (1.4b), and $N(k)$ is given by equation (2.16).

\hfill {\bf QED}

\section{Spectral Functions Associated with Different Lax Pairs}
Equation (2.14) expressed a ``symmetry'' relation for the eigenfunction $\MM(t,k)$ associated with
the second Lax pair (2.11).  The spectral functions $A(k)$ and $B(k)$ are defined in terms of the
eigenfunction $M(t,k)$ associated with the first Lax pair (2.9).  In what follows we present a
general result which, starting with the ``symmetry'' relation satisfied by $\MM(t,k)$, yields the
``symmetry'' relations satisfied by $\{ A(k),B(k)\}$.

\vskip .2in
\noindent {\bf Proposition 3.1}  Let $0<t<T$, $k \in {\mathbb{C}}$, $k
\neq 0$.  Let $M(t,k)$ be the unique solution of the $2\times 2$
matrix equations
$$ 
M_t = V(t,k)M, \quad V(t,k) = \tilde{Q}(0,t,k) - if_{2}(k) \sigma_{3},
$$
$$M(0,k) = \diag(1,1), \eqno (3.1)$$
where $\tilde{Q}, f_{2}, \sigma_{3}$ are as in Definition 2.1 (see
equation (2.1)).  Assume that $M(t,k)$ has the form (2.4).  Define the
spectral functions $A(k)$ and $B(k)$ by equations (2.5) where $f_2(k)$
satisfies the relation (2.6).

Let $\MM(t,k)$ be the unique solution of the $2 \times 2$ matrix equations
$$\MM_t = \V(t,k)\MM, $$
$$\MM(0,k) = \diag(1,1). \eqno (3.2)$$
Suppose that there exists a non-singular $2\times 2$ matrix $H(t)$ such that 
$$H_t(t) = \V(t,k)H(t)-H(t)V(t,k). \eqno (3.3)$$
Assume that $\MM(t,k)$ satisfies the ``symmetry'' relation
$$\MM(t,\nu(k)) = N(k)\MM(t,k)N(k)^{-1}, \eqno (3.4)$$
where $N(k)$ is a $2\times 2$ nonsingular matrix.

Then $M(t,k)$ satisfies the symmetry relation
$$M(t,\nu(k)) = F(t,k)M(t,k)F(0,k)^{-1}, \quad F(t,k) = H(t)^{-1}N(k)H(t). \eqno (3.5)$$
Furthermore, the spectral functions $A(k)$ and $B(k)$ satisfy the symmetry relation
$$ \curl{A(\nu(k))e^{-if_2(k)T}}{- B(\nu(k))e^{-if_2(k)T}}{ - \rho \overline{B(\nu(\bar
k)}e^{\overline{if_2(\bar k)}T}}{ \overline{A(\nu(\bar k)}e^{\overline{if_2(\bar k)}T}} = $$
$$ H(T)^{-1}N(k)H(T) \curl{A(k)e^{-if_2(k)T}}{- B(k)e^{-if_2(k)T}}{- \rho \overline{B(\bar
k)}e^{\overline{if_2(\bar k)}T}}{ \overline{A(\bar k)}e^{\overline{if_2(\bar k)}T}} H(0)^{-1}
N(k)^{-1}H(0). \eqno (3.6)$$

\noindent{\it Proof.}  If the $2\times 2$ matrices $\V(t,k)$ and $V(t,k)$ are related by equation
(3.3),  then the $2\times 2$ matrices $\MM(t,k)$ and $M(t,k)$ are related by the equation
$$ \MM(t,k) = H(t)M(t,k)H(0)^{-1}. \eqno (3.7)$$
Indeed, equation (3.7) is identically satisfied at $t=0$.  Furthermore replacing in equation
(3.2a) $\MM(t,k)$ by the rhs of equation (3.7) we find
$$H_tMH(0)^{-1} + HM_tH(0)^{-1} = \V HMH(0)^{-1}. $$
Replacing in this equation $M_t$ by $VM$, and using equation (3.3) we find an identity. 

Since $\MM(t,k)$ satisfies the ``symmetry'' relation (3.4), and $M(t,k)$ is related with $\MM(t,k)$
through equation (3.7), it is easy to show that $M(t,k)$ satisfies the symmetry relation (3.5).  Indeed,
this latter equation follows from (3.7) by replacing $k$ with $\nu(k)$ and then using equations
(3.4) and (3.7).

Having established the ``symmetry'' relation (3.5) it is straightforward to obtain a symmetry
relation for the spectral functions $ \{ A(k),B(k)\}$.  Indeed, evaluating equation (3.5) at $t=T$,
and expressing $M(T,k)$ in terms of the spectral functions (see equations (2.4), (2.5)), equation
(3.5) yields equation (3.6).  \hfill {\bf QED}

\paragraph{The sG Case}  For the sG equation
$$ f_2(k) = \quarter(k + \frac{1}{k}), $$
$$V(t,k) = Q(0,t,-k) - \frac{i}{4} (k+ \frac{1}{k})\sigma_3, \quad \V(t,k) = \V(0,t,k), \eqno
(3.8)$$
where $Q(x,t,k)$ and $\V(x,t,k)$ are defined by equations (2.10) and (2.12b) respectively.  It can
be verified that in this case $M(t,k)$ has the form (2.4) with $\rho =-1$.  Furthermore, equation
(3.3) is valid with the $2\times 2$ nonsingular matrix $H(t)$ given by
$$
H(t) = e^{-\frac{i\pi}{4}}
\left( \begin{array}{ll}
e^{\frac{iq(0,t)}{4}} & e^{\frac{iq(0,t)}{4}} \\ \\
-ie^{\frac{-iq(0,t)}{4}} & ie^{\frac{-iq(0,t)}{4}}
\end{array} \right). \eqno (3.9)
$$
Thus for the boundary condition (1.4b), the spectral functions $A(k)$ and $B(k)$ satisfy equation
(3.6) with $f_2(k)$, $H(t)$, and $N(k)$ given by equations (3.8a), (3.9), and (2.16) respectively.

\paragraph{Proof of Theorem 1.1}  

Let $q(x,t)$ be defined by equation ( 1.5) in terms of the
solution $\mu(x,t,k)$ of the $2\times 2$ matrix RH problem with the jump condition (1.6), where the jump
matrix $J$ is defined by equations (1.7) in terms of $\{ a(k),b(k), A(k),B(k)\}$.  Let $a(k)$ and
$b(k)$ be defined by equations (1.8) and (1.9).  Let $A(k)$ and $B(k)$ be defined by equations
(2.3)--(2.5) where $\tilde Q(0,t,k)$ in equation (2.3) equals $Q(0,t,-k)$ which is defined by
equation (2.10) with $q(0,t)$ and $q_x(0,t)$ replaced by $g_0(t)$ and $g_1(t)$.  It was shown in
[1] that if $g_0(t)$ and $g_1(t)$ are such that the global relation (1.1) is valid, then $q(x,t)$
satisfies the sG, and also $q(x,0) = q_0(x)$, $q_t(x,0) = q_1(t)$, $q(0,t) = g_0(t)$, $q_x(0,t) =
g_1(t)$.

Furthermore, it was shown in [1] that if $g_0(t) = \chi$ then the definition of $\{A(k),B(k)\}$
and the global relation (1.1) imply (1.10a).  Thus it only remains to be shown that if the
boundary condition (1.4b) is valid then $B/A$ satisfies equation (1.10b).  In this respect we note
that the definition of $\{ A(k),B(k)\}$ implies the symmetry relation (3.6).  In what follows we
show that this relation together with the global relation (1.1) imply
(1.10b).  For this purpose it is convenient to assume that $q(x,t) \to
0$ as $t \to \infty$ and to let $T \to \infty$.  In this case $A(k)$ and
$B(k)$ are \emph{not} entire functions but are analytic functions for
$k \in D_1 \cup D_3$.  Also, the global relation (1.1) becomes
$$
a(k)B(k) - b(k)A(k) = 0, \quad k \in D_1. \eqno (3.10)
$$
Note that $\{ A(k), B(k)\}$ are defined for $k \in D_1 \cup D_3$,
while $\{ a(k), b(k) \}$ are defined for $k \in D_1 \cup D_2$, thus
each term of equation (3.10) is well defined for $k \in D_1$.

The assumption $q(0,t) \to 0$ as $t \to \infty$, and the definitions
of $H(t)$ and $N(k)$, ie.~equations (3.9) and (2.16), imply
$$
H(\infty)^{-1} N(k) H(\infty) = \frac{1}{2}
\Bigg( 
\begin{array}{ll}
\alpha + \bar{\alpha} - 2i & \alpha - \bar{\alpha} \\
\alpha - \bar{\alpha} & \alpha + \bar{\alpha} + 2i
\end{array}
\Bigg), \eqno (3.11)
$$
where $\alpha(k)$ is defined by equation (2.21a).  Similarly
$$
H(0)^{-1} N(k)^{-1} H(0) = \frac{1}{2(1+\alpha\bar{\alpha})}
\Bigg( 
\begin{array}{ll}
\alpha + \bar{\alpha} + 2i\cos{(\frac{q_{0}(0)}{2})} & \bar{\alpha} -
\alpha - 2\sin{(\frac{q_{0}(0)}{2})}\\
\bar{\alpha} - \alpha + 2\sin{(\frac{q_{0}(0)}{2})} & \alpha +
\bar{\alpha} - 2i\cos{(\frac{q_{0}(0)}{2})}
\end{array}
\Bigg). \eqno (3.12)
$$
Using equations (3.11) and (3.12) in equation (3.6) (with $\rho =
-1$), solving for $A(1/k), B(1/k)$, and letting $T \to \infty$, we
find
\begin{equation*}
\begin{split}
A(\frac{1}{k}) &= \frac{\alpha + \bar{\alpha} -
  2i}{4(1+\alpha\bar{\alpha})} \Big\{ [\alpha + \bar{\alpha} +
  2i\cos{(\frac{q_{0}(0)}{2})}]A(k) - [\bar{\alpha} - \alpha +
  2\sin{(\frac{q_{0}(0)}{2})}]B(k) \Big\}, \\
B(\frac{1}{k}) &= \frac{\alpha + \bar{\alpha} -
  2i}{4(1+\alpha\bar{\alpha})} \Big\{ [\alpha + \bar{\alpha} -
  2i\cos{(\frac{q_{0}(0)}{2})}]B(k) - [\bar{\alpha} - \alpha -
  2\sin{(\frac{q_{0}(0)}{2})}]A(k) \Big\},
\end{split} \tag{3.13}
\end{equation*}
where $k \in D_1 \cup D_3$.

The ``symmetry'' relations (3.13) together with the global relation
(3.10), yield $B(k)/A(k)$ in terms of $\{a(k),b(k)\}$.  Indeed, for $k
\in D_1$,
$$
\frac{B(k)}{A(k)} = \frac{b(k)}{a(k)}, \quad k \in D_1.
$$
Replacing $k$ by $1/k$ in this equation, using equations (3.13), and
solving the resulting equation for $B(k)/A(k)$ we find equation
(1.10b).

It was shown in [1] that the solution of the basic RH problem is
independent of $T$.  Thus although the basic formula (1.10b) was
derived under the assumption that $q(x,t) \to 0$ as $t \to \infty$,
this formula is valid even without this assumption. \hfill {\bf QED}

\section{The Residue Conditions}

We \emph{assume} that $a(k)$ has $n$ simple zeros $\{ k_{j}
\}_{1}^{n}$, $n_{1}$ of which are in $D_{1}$ and the remaining
$n-n_{1}$ of which are in $D_{2}$.

We denote by $\dot{a}(k)$ the derivative of $a(k)$ with respect to
$k$, and we denote by $[\mu]_{1}$ and $[\mu]_{2}$ the first and second
columns of the $2\times2$ matrix $\mu$.

Let $\{ \lambda_{j} \}_{1}^{\Lambda}$ denote the zeros of $\Delta(k)$
for $k \in D_{2}$, where $\Delta(k)$ is defined by equation (1.12).

The following residue conditions are valid:
\begin{align*}
\Res_{k_{j}} [\mu(x,t,k)]_{1} &=
\frac{e^{2i\theta(k_{j})}}{\dot{a}(k_{j})b(k_{j})} [\mu(x,t,k_{j})]_{2},
\quad j = 1,\dots,n_{1}, \quad k_{j} \in D_{1}, \\
\Res_{\bar{k}_{j}} [\mu(x,t,k)]_{2} &=
-\frac{e^{-2i\theta(\bar{k}_{j})}}{\overline{\dot{a}(k_{j})}\,
  \overline{b(k_{j})}} [\mu(x,t,\bar{k}_{j})]_{1}, \quad j =
1,\dots,n_{1}, \quad \bar{k}_{j} \in D_{4}, \\
\Res_{\lambda_{j}} [\mu(x,t,k)]_{1} &=
-\frac{\overline{N(\bar{\lambda}_{j})}
  e^{2i\theta(\lambda_{j})}}{\dot{a}(\lambda_{j}) \dot{\Delta}(\lambda_{j})} [\mu(x,t,\lambda_{j})]_{2},
\quad j = 1,\dots,\Lambda, \quad \lambda_{j} \in D_{2}, \\
\Res_{\bar{\lambda}_{j}} [\mu(x,t,k)]_{2} &=
-\frac{N(\lambda_{j})
  e^{-2i\theta(\bar{\lambda}_{j})}}{
  \overline{\dot{a}(\lambda_{j})}\,
  \overline{\dot{\Delta}(\lambda_{j})}}
[\mu(x,t,\bar{\lambda}_{j})]_{1}, \quad j = 1,\dots,\Lambda, \quad
\bar{\lambda}_{j} \in D_{3}, \tag{4.1}
\end{align*}
where $N(k)$ denotes the numerators of equations (1.10), and $a(k),
b(k), \theta(x,t,k)$ are defined in Theorem 1.1.

In order to derive these equations we first recall that the matrix
$\mu(x,t,k)$ appearing in the RH problem of Theorem 1.1 is constructed
from three appropriate matrix solutions, $\{ \mu_{j}(x,t,k)
\}_{1}^{3}$, of the Lax pair (2.9).  These matrices can be written in
the column vector form
$$
\mu_{1} = (\mu_{1}^{(2)},\mu_{1}^{(3)}), \quad \mu_{2} =
(\mu_{2}^{(1)},\mu_{2}^{(4)}), \quad \mu_{3} =
(\mu_{3}^{(34)},\mu_{3}^{(12)}), \eqno (4.2)
$$
where superscripts denote the domains that these column vectors are
bounded and analytic ($\mu_{1}^{(2)}$ in $D_{2}$, $\mu_{3}^{(34)}$ in
$D_{3} \cup D_{4}$ etc.).  It is shown in [1] that the matrix $\mu$
has the following form:
\begin{align*}
\mu_{+} &= \left( \frac{\mu_{2}^{(1)}}{a(k)}, \mu_{3}^{(12)} \right)
\quad k \in D_{1};& \mu_{-} &= \left( \frac{\mu_{1}^{(2)}}{d(k)},
\mu_{3}^{(12)} \right), \quad k \in D_{2};& \\
\mu_{+} &= \left( \mu_{3}^{(34)},
\frac{\mu_{1}^{(3)}}{\overline{d(\bar{k})}} \right) \quad k \in
D_{3};& \mu_{-} &= \left( \mu_{3}^{(34)},
\frac{\mu_{2}^{(4)}}{\overline{a(\bar{k})}} \right), \quad k \in
D_{4};& \tag{4.3}
\end{align*}
where
$$
d(k) = a(k) \overline{A(\bar{k})} + b(k) \overline{B(\bar{k})}. \eqno (4.4)
$$

In order to derive equation (4.1a) we condsider the equation $\mu_{-}
= \mu_{+} J_{4}$, where $J_{4}$ is defined in Theorem 1.1 and
$\mu_{+}, \mu_{-}$ are given by the first and fourth equations in
(4.3).  The second column of this equation yields
$$
\mu_{3}^{(12)} = a \mu_{2}^{(4)} + b e^{-2i\theta} \mu_{2}^{(1)}.
$$
Evaluating this equation at $k_{j}$ we find
$$
\mu_{3}^{(12)}(k_{j}) = b(k_{j}) e^{-2i\theta(k_{j})}
\mu_{2}^{(1)}(k_{j}),
$$
where for simplicity of notation we suppress the $(x,t)$ dependence of
$\mu_{3}^{(12)}$ and $\mu_{2}^{(1)}$.  Hence
$$
\Res_{k_{j}} [\mu]_{1} = \frac{\mu_{2}^{(1)}(k_{j})}{\dot{a}(k_{j})} =
\frac{e^{2i\theta(k_{j})}}{\dot{a}(k_{j})b(k_{j})}
  \mu_{3}^{(12)}(k_{j}),
$$
and since $\mu_{3}^{(12)} = [\mu]_{2}$, equation (4.1a) follows.  The
derivation of equation (4.1b) is similar.

In order to derive equation (4.1c) we consider the equation $\mu_{-} =
\mu_{+} J_{1}$, where $J_{1}$ is defined in Theorem 1.1 and $\mu_{+},
\mu_{-}$ are given by the first and second equations in (4.3).  The
first column of this equation yields
$$
\frac{\mu_{1}^{(2)}}{d(k)} = \frac{\mu_{2}^{(1)}}{a(k)} - \frac{
  \overline{N(\bar{k})} e^{2i\theta}}{ a(k) [ a(k)
    \overline{D(\bar{k})} + b(k) N(\bar{k}) ]} \mu_{3}^{(12)}. \eqno
  (4.5)
$$
Following arguments similar to those used in [3] it can be shown that
the zeros of $d(k)$ in $D_{2}$ coincide with the zeros of $\Delta(k)$
in $D_{2}$, thus equation (4.5) yields
$$
\mu_{1}^{(2)}(\lambda_{j}) = - \frac{\overline{N(\bar{\lambda}_{j})}
  e^{2i\theta(\lambda_{j})}}{a(\lambda_{j})}
\mu_{3}^{(12)}(\lambda_{j}).
$$
Hence
$$
\Res_{\lambda_{j}} [\mu]_{1} =
\frac{\mu_{1}^{(2)}}{\dot{d}(\lambda_{j})} = -
\frac{\overline{N(\bar{\lambda}_{j})}
  e^{2i\theta(\lambda_{j})}}{a(\lambda_{j}) \dot{\Delta}(\lambda_{j})}
\mu_{3}^{(12)}(\lambda_{j}),
$$
and since $\mu_{3}^{(12)} = [\mu]_{2}$, equation (4.1c) follows.  The
derivation of (4.1d) is similar.

\section{Conclusion}

It was shown in [1] and [3] that there exists a particular class of boundary conditions for which
initial-boundary value problems on the half-line can be solved with the same level of efficiency
as the classical initial value problem on the line.  These ``linearizable'' boundary conditions were
identified by the requirement that the eigenfunction $M(t,k)$, which satisfies the $t$-part of the
associated Lax pair evaluated at $x=0$ (as well as the condition that $M(0,k)$ equals the identity
matrix), satisfies the ``symmetry'' relation
$$M(t,\nu(k)) = N(k)M(t,k)N(k)^{-1}; \eqno (5.1)$$
in this equation the map $k\rightarrow \nu(k)$ is the map which leaves the dispersion relation of
the linearized version of the given nonlinear PDE invariant (for the sG, $\nu(k) = 1/k$) and $N(k)$
is a non-singular matrix.

In this paper we have generalized equation (5.1) to the equation
$$M(t,\nu(k)) = F(t,k) M(t,k)F(0,k)^{-1}, \eqno (5.2)$$
where $F(t,k)$ is a nonsingular matrix. Furthermore we have given an algorithmic way of
constructing $F$ by making use of the existence of different Lax pair formulations for the same
nonlinear PDE.  We expect that the systematic use of B\"acklund transformations will provide an
approach to computing $F(t,k)$ for any linearizable boundary
condition.

The main advantage of our method is {\it not} that it identifies linearizable boundary conditions,
but that for such boundary conditions it expresses the solution
$q(x,t)$ in terms of a simple
Riemann-Hilbert (RH) problem.  The basic features of this problem are similar with the basic
features of the RH problem characterizing the solution of the classical initial-value problem,
namely these two RH problems: (a) Have the {\it same} explicit $(x,t)$ dependence.  (b) Their jump matrices
involve the functions $a(k)$ and $b(k)$ which are constructed from the initial
conditions in a similar manner.  Regarding differences between these
two RH problems, we note that the RH problem associated with initial-boundary value problems
has the novelty that it is formulated on a more complicated contour, and it also involves
some additional jump functions which however can be {\it explicitly} written in terms of $a(k)$
and $b(k)$ (see equations (1.10)).  The existence of a more complicated contour does {\it not} add
any significant complexity to the analysis of the RH problem.  Also in some cases it is possible
to map this contour to the usual contour which is the real $k$-axis.  This is actually the case
for the sG and the nonlinear \s [7] (but {\it not} for the Korteweg-deVries and the modified
Korteweg-deVries).

The simplicity of the basic RH problem has important implications for the analysis of the long
time asymptotics.  Indeed, in the case that the boundary conditions decay as $t\rightarrow
\infty$, it is possible using the Deift-Zhou approach [8] to obtain a complete characterization of
the solution as $t\rightarrow \infty$ and $x/t=O(1)$. The general structure of the asymptotics is
given in [9], where it is shown that the solution is dominated by solitons. A detailed
investigation of these solitons for the boundary condition (1.4b), as well as for the boundary
condition
$$ q(0,t) = \chi, \quad 0<t<t_0; \quad q(0,t) =0, \quad t_0<t<\infty,$$
will be presented elsewhere. We also note that the simplicity of the basic RH problem makes it
possible, using the Deift-Zhou-Venakides approach [10], to study the zero dispersion limit of
initial-boundary value problems [7], [11].

Several authors have identified linearisable boundary conditions using
the existence of symmetries and conservation laws, see for example
[12]--[13].  The analysis of such boundary conditions using several
formal RH problems was presented in [14].  The particular cases of
either $\chi_{1}=0$ or $\chi_{2}=0$ are discussed in [15] using an
extension of the problem from the half line to the infinite line (such
an extension is \emph{not} possible for PDEs with a third order
derivative such as the KdV and the modified KdV equations).  A discussion of the
physical significance of the sG with the boundary condition (1.4b) as
well as several approaches for the analysis of this problem can be
found in [16]--[21].  In particular in [18] the question of the
integrability of both the classical and quantum sine-Gordon theory
involving $m^{2} \sin{(\beta q(x,t))}/\beta$ with the boundary condition
$$
\frac{\partial{q(0,t)}}{\partial{x}} +
\frac{\partial{}V(q(0,t))}{\partial{q}} = 0, \eqno (5.3)
$$
was investigated.  By demanding that a modification of the first
non-trivial integral of motion of the usual theory remain conserved it
was found that in general
$$
V(q(0,t)) = \frac{Am}{\beta} \cos{ \left( \beta \frac{(q(0,t) -
    q_{0})}{2} \right)}, \eqno (5.4)
$$
where $A$ and $q_0$ are arbitrary real constants.  In [18] the
existence of an infinite number of integrals of motion for this system
was assumed and the associated quantum field theory was studied.  In
[17] the question of integrability of the classical system with the
boundary condition (5.3) was addressed.  The results of [16] were used
to prove that an infinite number of integrals of motion do exist, but
only for certain $V(q)$.  It was found that (5.4) is the most general
boundary term compatible with the existence of infinitely many
conserved quantities.  This result thus agrees
with [18].  The boundary condition (5.4) with $q_0=0$ or $q_0=\pi/2$
has appeared in classical considerations of sine-Gordon theory.  It
was suggested in [18] that the scattering theory can depend on the
extra parameter $q_0$; a similar question was investigated in [20] for
real coupling affine Toda field theory.

\paragraph{Acknowledgments}
I am deeply grateful to E. Corrigan for suggesting this problem to me,
and to A.R. Its for some important suggestions and in particular for
showing me the relation between the two different Lax pair
formulations of the sG equation.

\end{document}